\documentclass[aps,prl, preprint, showpacs, nobibnotes,12pt]{revtex4-1}
\usepackage{graphicx}
\usepackage{nomencl}
\usepackage{amssymb,amsmath}
\usepackage{subfig}
\usepackage{wasysym}

\begin{document}














\vspace{1cm}
\centerline{\large{On the Instability and Critical Damping Conditions, $k\tau = 1/e$ and $k\tau = \pi/2$, }}
\centerline{\large{ of the Equation $\dot{\theta} = -k \theta(t-\tau)$}}

\vspace{0.3cm}

\centerline{{\small Z. Jane Wang}}

\vspace{0.3cm}

\centerline{jane.wang@cornell.edu}
\centerline{{\small Date: March 16, 2014}}
\vspace{0.5cm}

In this note,  I show that it is possible to use elementary mathematics, instead of the machinery of Lambert function, Laplace Transform,  or numerics,  to derive the instability condition, $k \tau = \pi/2$, and the critical damping condition, $k\tau = 1/e$, for the time-delayed equation $\dot{\theta} = -k \theta(t-\tau)$.   I hope it will be useful for the new comers  to this equation, and perhaps even to the experts if this is a simpler method compared to other versions.

I had assumed that these are textbook results, but after asking around, it doesn't appear to be the case. The results $k\tau=\pi/2$\cite{Milton2011}  $k\tau=1/e$\cite{LandCollett1974} were stated in the literature, though without derivations.  The current research on time-delayed differential equation  typically starts  by expressing the general solution in terms of infinitely series.   This does not lead to obvious explicit solutions, such as  these two conditions.  Therefore this note is in a way the first leg of `six degrees of separation' experiment.  Except that instead of sending the question to my best guesses of who might know the source, or search online, or go to the library,  it  will wander on the web.  Perhaps we can compare the speed of these different search methods.  I will try this one first.

To solve for $\dot{\theta} = -k \theta(t-\tau)$, we start by writing $\theta = e^{\lambda t}$, where $\lambda$ is a complex number, $\lambda = a + ib$.  This leads to the characteristic equation for $\lambda$:

\begin{eqnarray}
\lambda =&- k e ^{-\lambda \tau}
\label{eq-lambda}
\end{eqnarray}

\noindent  Because of the time delay $\tau$,  the characteristic equation, Eq. (\ref{eq-lambda})  is a transcendental equation and does not admit explicit solutions. It is a special case of Lambert function
\cite{Richard}.  The rest of this note is to show how to  find the instability and critical conditions without manipulating the Lambert function or using the explicit values of $\tau$ and $\lambda$.

\section{ Stability Condition  $k\tau < \pi/2$} 

The solution $e^{\lambda t} = e^{a+ib}$ becomes unstable when $a>0$. We will see that this occurs when  $k\tau \ge \pi/2$.  According to eq.(\ref{eq-lambda}),  $a + ib = - ke^{-(a+ib)\tau}$. Separating the real and imaginary parts, we have

\begin{eqnarray}
a &=& -k e^{-a\tau} \cos{b\tau} \\
b &=& ke^{-a\tau}\sin{b\tau}
\label{eq-egv}
\end{eqnarray}

\noindent in other words, 

\begin{eqnarray}
a^2+b^2 &=& k^2 e^{-2a\tau} \label{eq-ab1} \\
-\frac{b}{a} &=&  \tan{b\tau} \label{eq-ab2}
\label{eq-egv}
\end{eqnarray}

\begin{figure}[h] 
   \centering
   \includegraphics[width=3in]{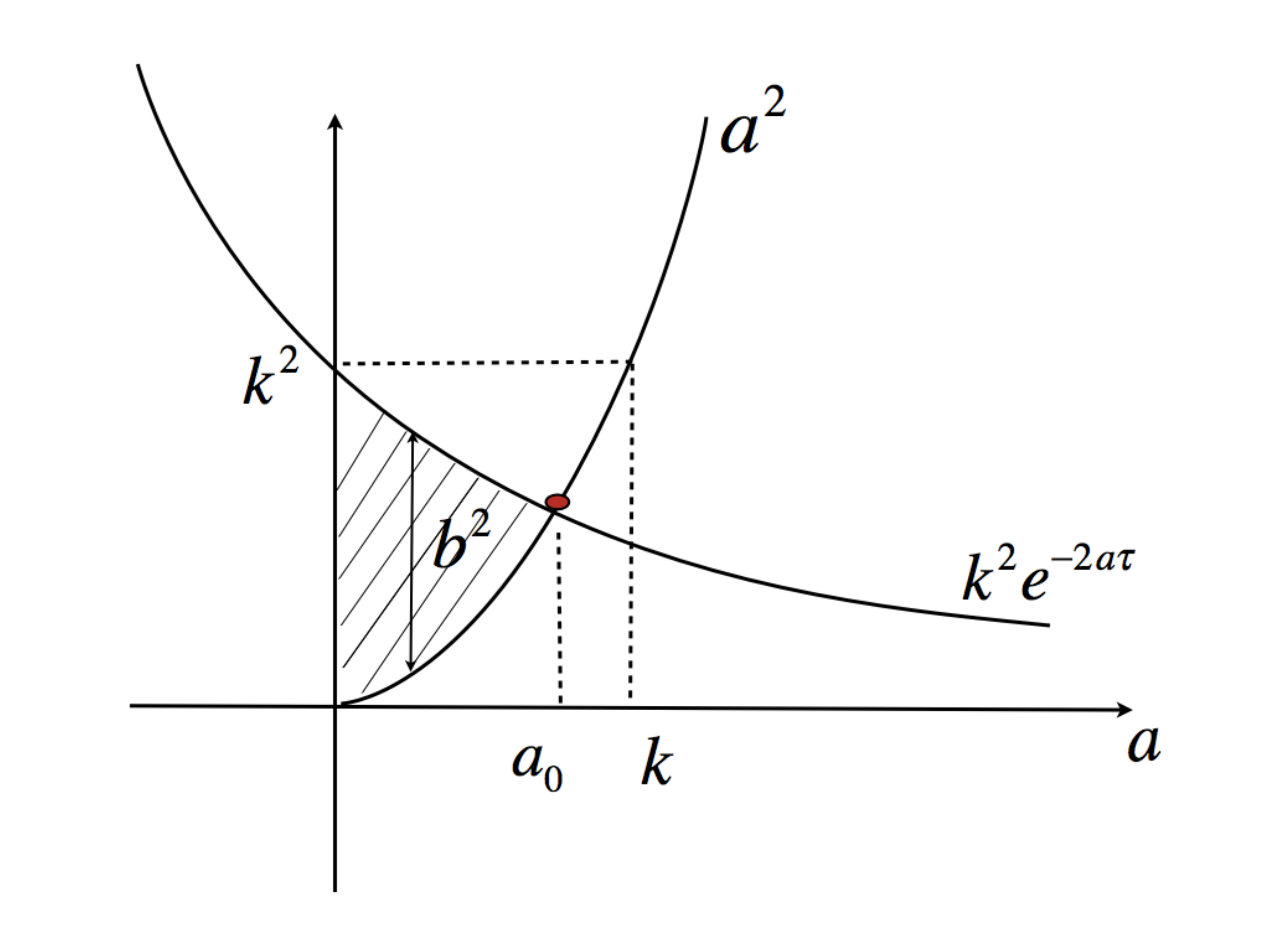} 
   \includegraphics[width=3in]{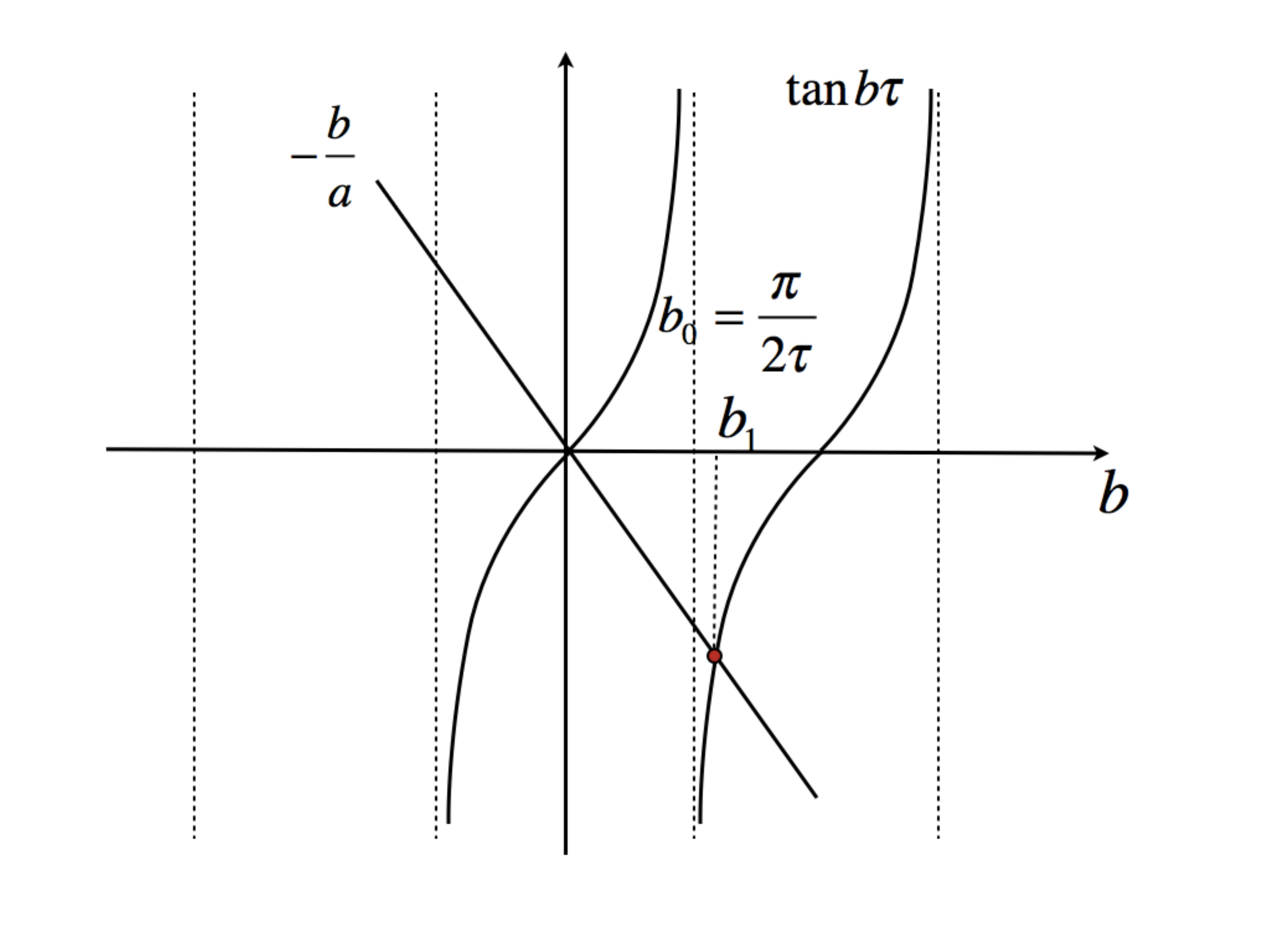} 
  \caption{Finding roots to $a^2+b^2 = k^2 e^{-2a\tau}$  and  $-\frac{b}{a} =  \tan{b\tau}$}
   \label{fig-instability}
\end{figure}

\noindent The question is when do they admit solutions with $a>0$?  First consider eq.~(\ref{eq-ab1}). If we plot $k^2 e^{-2a\tau}$ and  $a^2$ against $a$, we see that the intersection occurs at $a_0 <k$ (Fig.~\ref{fig-instability}, left).  Since $b^2 =k^2 e^{-2a\tau} - a^2$, which  corresponds to the shaded area,  it can be seen that $b^2 <k^2$, or $|b| <k$.  Now consider eq.~(\ref{eq-ab2}) and plot  $-\frac{b}{a}$ and $\tan{b\tau}$ against $b$.   The first branch of  $\tan{b\tau}$ is within $[-b_0, b_0]$, with $b_0 = \frac{\pi}{2\tau}$. The line $-\frac{b}{a}$ intersects $\tan {b\tau}$ at $b_1 > b_0 = \frac{\pi}{2\tau}$. 

Combining the two conditions, $|b| < k$ and $b_1 >  \frac{\pi}{2\tau}$, we have $k > b_1 >  \frac{\pi}{2\tau}$. This inequality gives $k\tau> \frac{\pi}{2}$,  a necessary condition for instability. 

From the same graph, we can also see that $a_0$ has an upper-bound $k$, which means that the unstable solutions grows at most as $e^{k\tau}$.

\section{Critical Damping Condition $k\tau = 1/e$}

 The graph method turns out to be helpful in deriving the critical damping condition $k\tau = 1/e$, which showed up in fly's\cite{LandCollett1974} and  beetle's\cite{HaselsteinerGibertWang2014} pursuit dynamics. 
Among the damping solutions, $e^{at}$, we can show that the fastest decay mode occurs at $k\tau=1/e$.  For damping solutions, $\lambda=a<0$,  the characteristic equation becomes  $a = -k e ^{-a \tau}$.  Multiply $\tau$ on both sides,  $a \tau = -k \tau e ^{-a \tau}$, and let $x=a\tau$, we have 

\begin{equation}
xe^{x} = - k\tau
\label{eq-x}
\end{equation}

To find solutions to eq.(\ref{eq-x}), let's plot $xe^{x}$ against $x$.  The solutions of $a$ are at the intersection of this curve and $-k\tau$.   In general, there are a pair of intersection points,  $-a_1$ and $-a_2$, which become degenerate at  the minimum of the curve. The general solution is therefore the linear superposition of the two, $A e^{-a_1t} + Be^{-a_2t}$.  For $a_1 < a_2$, the solution is dominated by $Ae^{-a_1t}$. 

\begin{figure}[h] 
   \centering
   \includegraphics[width=4in]{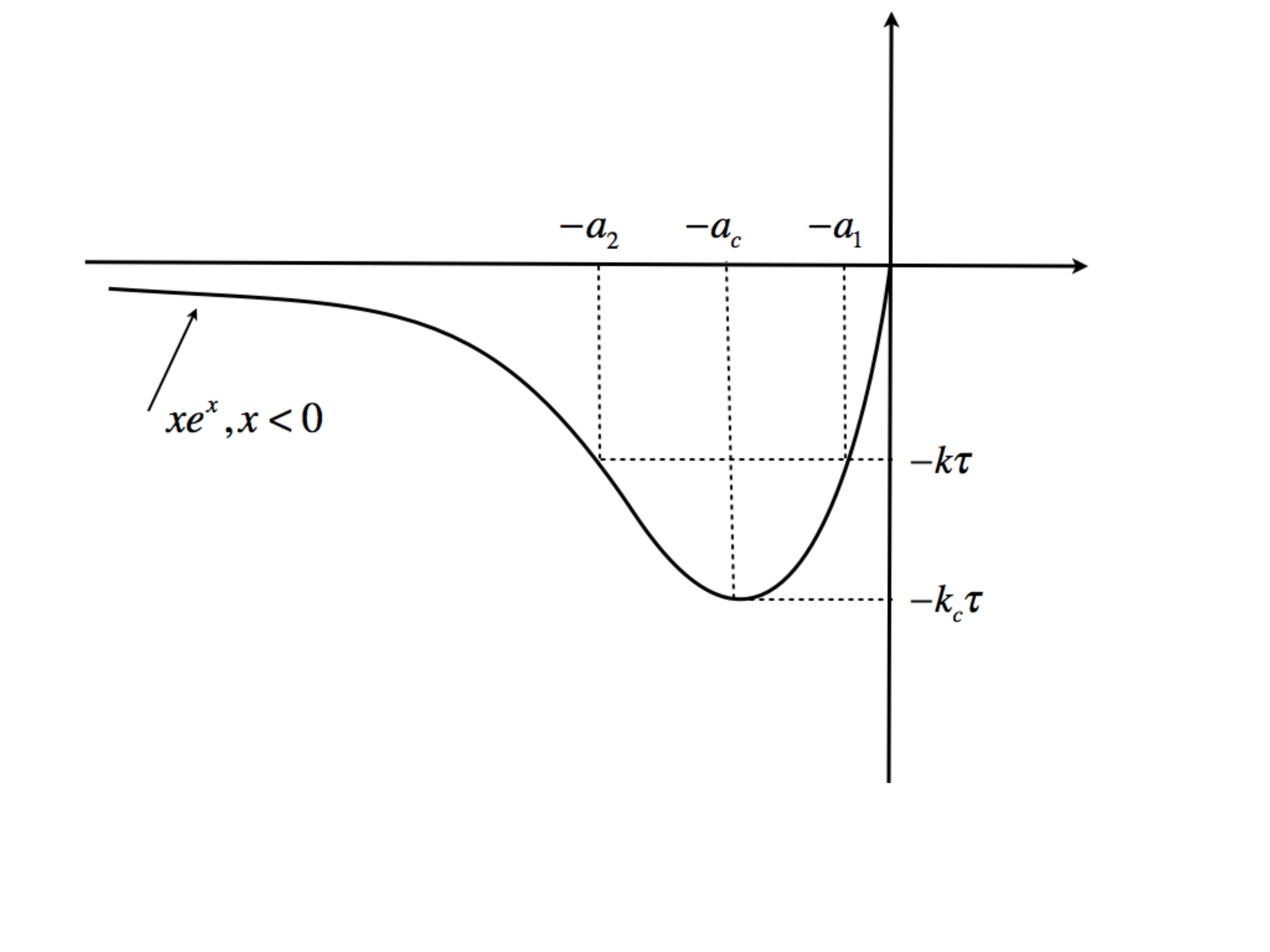} 
  \caption{The solutions to $xe^{x} = - k\tau$}
   \label{fig-damping}
\end{figure}

By definition, if the solution is of the form of $e^{-at}$, the fastest decay occurs at the maximum of all possible $a$. This is equivalent to finding  the maximum of all $a_1$.   Fig. \ref{fig-damping} shows that it occurs at the bottom of the curve of $y=xe^x$.

This turns the problem of finding the minimum of $a_1$ into an easier problem of determining the minimum of $y(x)=xe^x$.   The minimum occurs when $dy/dx=0$, $e^x +x e^x =0$,  which gives  $x_c=-1$. 

Going back to the definition of $x = a\tau$,   we have the critical damping condition, $a_c=-1/\tau$.  Recall that $a_c$ also satisfies the characteristic equation,  $a_c \tau = -k \tau e ^{-a_c \tau}$, we can express  the critical condition in the sought form, in terms of  $k$ and $\tau$,

\begin{equation}
k\tau = 1/e
\end{equation}

{}

\end{document}